# AXIOM: Advanced X-ray Imaging Of the Magnetosphere

G. Branduardi-Raymont[1*], S. F. Sembay[2], J. P. Eastwood[3], D. G. Sibeck[4], A. Abbey[2], P. Brown[3], J. A. Carter[2], C. M. Carr[3], C. Forsyth[1], D. Kataria[1], S. Kemble[5], S. E. Milan[2], C. J. Owen[1], L. Peacocke[5], A. M. Read[2], A. J. Coates[1], M. R. Collier[4], S. W. H. Cowley[2], A. N. Fazakerley[1], G. W. Fraser[2], G. H. Jones[1], R. Lallement[6], M. Lester[2], F. S. Porter[4], T. K. Yeoman[2]

[1] *University College London, Mullard Space Science Laboratory, Holmbury St Mary, Dorking, Surrey, UK*

[2] *Department of Physics and Astronomy, University of Leicester, Leicester, UK*

[3] *Blackett Laboratory, Imperial College London, London, UK*

[4] *NASA Goddard Space Flight Center, Greenbelt, MD, USA*

[5] *Astrium Ltd, Stevenage, UK*

[6] *LATMOS/Institute Pierre Simon Laplace, Paris, France*

**Abstract** Planetary plasma and magnetic field environments can be studied in two complementary ways – by in situ measurements, or by remote sensing. While the former provide precise information about plasma behaviour, instabilities and dynamics on local scales, the latter offers the global view necessary to understand the overall interaction of the magnetospheric plasma with the solar wind. Some parts of the Earth's magnetosphere have been remotely sensed, but the majority remains unexplored by this type of measurements. Here we propose a novel and more elegant approach employing remote X-ray imaging techniques, which are now possible thanks to the relatively recent discovery of solar wind charge exchange X-ray emissions in the vicinity of the Earth's magnetosphere. In this article we describe how an appropriately designed and located X-ray telescope, supported by simultaneous in situ measurements of the solar wind, can be used to image the dayside magnetosphere, magnetosheath and bow shock, with a temporal and spatial resolution sufficient to address several key outstanding questions concerning how the solar wind interacts with the Earth's magnetosphere on a global level. Global images of the dayside magnetospheric boundaries require vantage points well outside the magnetosphere. Our studies have led us to propose 'AXIOM: Advanced X-ray Imaging Of the Magnetosphere', a concept mission using a Vega launcher with a LISA Pathfinder-type Propulsion Module to place the spacecraft in a Lissajous orbit around the Earth – Moon L1 point. The model payload consists of an X-ray Wide Field Imager, capable of both imaging and spectroscopy, and an in situ plasma and magnetic field measurement package. This package comprises a Proton-Alpha Sensor, designed to measure the bulk properties of the solar wind, an Ion Composition Analyser, to characterise the minor ion populations in the solar wind that cause charge exchange emission, and a Magnetometer, designed to measure the strength and direction of the solar wind magnetic field. We also show simulations that demonstrate how the proposed X-ray telescope design is capable of imaging the predicted emission from the dayside magnetosphere with the sensitivity and cadence required to achieve the science goals of the mission.[+]

**Keywords** X-rays. Space telescope. Space plasma instrumentation. Magnetometer. Techniques: Imaging, spectroscopy, plasma and field analysers.

# 1 Introduction

Over the past 50 years, in situ observations of the Earth's magnetosphere have provided very important data which have been used to characterise the basic physics that controls the plasma interaction between the Earth and the Sun. However, in situ observations are fundamentally limited by the number of available spacecraft. To overcome this limitation, ultimately a new approach is necessary. In particular, the experience of auroral and inner magnetospheric physics teaches us that imaging can provide the global view that is needed to understand the overall interaction between the solar wind and the magnetosphere. The question then becomes one of how to image the magnetosphere, and in particular the boundaries where it interacts with the solar wind. The answer arises from the relatively recent discovery of solar wind charge exchange (SWCX) X-ray emission.

## 1.1 The charge exchange (CX) process

As highlighted in the recent comprehensive review by Dennerl (2010, and references therein) charge exchange (CX), or charge transfer, has been investigated since the dawn of atomic physics as a mechanism leading to electromagnetic radiation. Its high efficiency in producing X-ray emission, though, was clearly understood only some 15 years ago. Basically the CX process involves the encounter of an ion with a neutral atom or a molecule, as a consequence of which the ion acquires an electron and is left in an excited state from which it decays releasing radiation of a well defined wavelength. Recognition that if the initial charge of the ions is high enough, X-rays are efficiently produced came with the discovery of cometary X-ray emission, and its interpretation as the result of CX (Lisse et al. 1996, Cravens 1997); in this case highly charged ions of the solar wind undergo CX with neutrals in the cometary coma in what is commonly named solar wind charge exchange, or SWCX.

Fig. 1 shows a cartoon of such a process (adapted from Dennerl 2009): A H-like oxygen ion picks up an electron from a water molecule in the comet's coma, turns into an excited He-like ion and subsequently decays emitting a characteristic X-ray line at 561 eV. Since the initial discovery, much theoretical and laboratory work has been dedicated to investigating CX; this process is now recognised as ubiquitous in the Universe, and as playing a very important role in the X-ray emission of sources as diverse as planets, including the geocorona, the heliosphere, stellar winds, the interstellar medium, galaxies and clusters of galaxies.

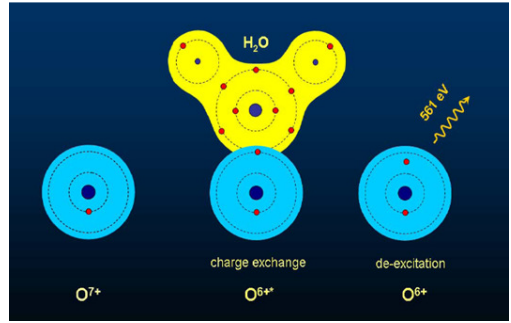

**Fig. 1** Cartoon illustrating the CX process between a H-like O ion and a cometary water molecule (adapted from Dennerl 2009)

An interesting consequence of CX is that if the collisions produce energetic neutral atoms (ENA) these are no longer bound to the local magnetic field, and thus can travel far maintaining a memory of their initial energy and direction (e.g. Hsieh et al. 1992). ENA detection thus provides a global picture of the plasma distribution, just like observations of X-ray emissions generated by ions in much higher charge states.

Directly relevant to the approach taken in this paper was the discovery of variable CX spectral signatures in Chandra observations of the dark Moon (Wargelin et al. 2004) which indicated that SWCX is taking place in the Earth's exosphere, i.e. in the very tenuous outer layers of the Earth's atmosphere, best known for their 'geocoronal' Lyα emission observed to extend to ~15 Earth radii ($R_E$; see Østgaard et al. 2003 for terrestrial high altitude hydrogen density profiles, and references therein for earlier exospheric models). Interactions of solar wind ions with the Earth's exospheric neutrals produce characteristic soft X-ray lines, with intensities that peak in the cusps and magnetosheath, two regions where both solar wind and neutral exospheric densities are high. Observations from XMM-Newton confirm that the emissions peak in the subsolar magnetosheath and track solar wind variations over a wide range of densities (Snowden et al. 2004, 2009; Carter et al. 2011).

In summary, the observations described above indicate that SWCX can be used to image the boundaries that form on the dayside of the Earth's magnetosphere.

## 1.2 A novel approach to imaging techniques

Several techniques for imaging the magnetosphere have already been employed: they include Energetic Neutral Atoms (ENA) imaging (e.g. Roelof 1989, Brandt et al. 2002, Mitchell et al. 2003), radio imaging (Green and Reinisch 2003), and $He^+$ EUV imaging (Sandel et al. 2003). However, there are only two ways to image the Earth's foreshock, magnetosheath, magnetopause and cusps (see sec. 2.1 for a description of the solar wind – magnetosphere interaction): ENA and SWCX soft X-rays. The two processes are complementary, since charge exchange underpins both. An advantage of using SWCX X-rays is that there is little or no contribution by the same mechanism from processes within the magnetosphere; so soft X-rays allow observations to focus on the dayside interactions processes. Moreover, current ENA imaging techniques (e.g. by IBEX) require hours to image the entire dayside interaction region (Fuselier et al. 2010), whereas the relevant interaction processes take place on timescales from one min to an hour. So a real step change is required if we are to improve substantially our knowledge of solar-terrestrial relationships, and this is offered by imaging in SWCX X-rays.

A wide-field-of-view soft X-ray telescope with spectroscopic capabilities located outside the Earth's magnetosphere can determine the spatial extent and track the motion of global and mesoscale boundary structures, evaluate the roles played by individual ion species, and thereby connect magnetospheric responses to solar wind drivers. Coupled with plasma and magnetic field instrumentation for simultaneous measurements of the input solar wind parameters, such an X-ray telescope will be a powerful, novel tool for both fully validating existing models and providing the new data required to develop a more accurate understanding of the solar wind – magnetosphere interaction. This is the motivation behind the 'AXIOM: Advanced X-ray Imaging Of the Magnetosphere' concept mission, which has been developed in response to the European Space Agency (ESA) call for medium (M3) mission proposals in 2010.

# 2 Scientific objectives and requirements

## 2.1 The dayside solar wind – magnetosphere interaction

The Earth's magnetic field carves out a cavity known as the magnetosphere in the collisionless supersonic and super-Alfvénic solar wind plasma. The solar wind flow compresses the sunward side of the magnetosphere but stretches the nightside out into a long magnetotail. A relatively sharp transition from hot tenuous magnetospheric plasmas to colder, denser, shocked solar wind plasmas marks the outermost boundary of the magnetosphere, known as the magnetopause. Cusp indentations at high latitudes on the dayside magnetopause denote locations where field lines divide to close in the opposite hemisphere or in the solar wind/distant magnetotail. As illustrated in Fig. 2, interconnected solar wind/magnetospheric magnetic field lines within the cusps enable solar wind plasma to penetrate deep into the magnetosphere, all the way to the ionosphere.

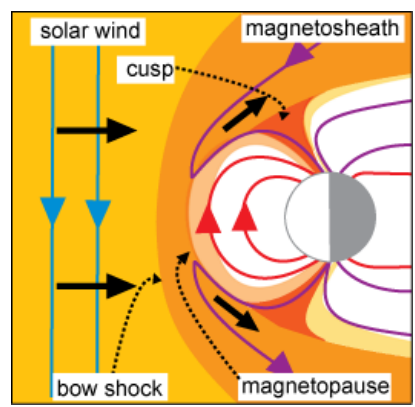

Because the solar wind is supersonic, a collisionless bow shock forms upstream of the magnetopause. The manner in which the solar wind is slowed, compressed, heated and diverted at planetary bow shocks is of key importance in understanding how the solar wind interacts with planetary magnetospheres, since it is the shocked magnetosheath plasma, not the pristine solar wind, that interacts with the magnetosphere at the magnetopause.

The position and shape of the magnetopause change continually as the Earth's magnetosphere responds to constantly varying solar wind dynamic pressures. The magnetosheath plasma flow depends on the shape and location of the shock, which in turn depends on the shape and location of the magnetopause, making an analytic solution intractable. Processes within the magnetopause can also alter its position and shape. Periods of strong magnetic shear across the subsolar magnetopause favour magnetic reconnection.

**Fig. 2** Sketch of the dayside magnetosphere: The magnetopause represents the outer boundary of the magnetosphere, and is compressed on the dayside. Solar wind is heated and deflected at the bow shock to flow around the magnetosphere.

Reconnection enables plasma to flow antisunward through the magnetopause boundaries, the cusps, and over the polar caps, setting up current systems within the magnetosphere that cause the subsolar magnetopause to erode towards the planet. The addition of magnetic flux to the nightside tends to increase the size of the tail and thus the degree of 'flaring' away from the magnetotail axis. Reconnection on the dayside magnetopause is therefore thought to cause the shape of the magnetopause to become blunter. By contrast, variations in the solar wind dynamic pressure should cause self-similar changes in magnetospheric dimensions.

While the Polar and IMAGE missions have provided global imaging of the Earth's aurorae and inner magnetosphere, most of our present knowledge of the dayside magnetosphere and its interaction with the solar wind has been derived from in situ observations. Historically, such measurements have returned precise information about local plasma dynamics and microphysics, but they do not provide the global view needed for a full understanding of the solar wind – magnetosphere interaction. In the absence of a flotilla of spacecraft, this requires global imaging. More recently, the dayside magnetospheric measurements obtained by Fuselier et al. (2010) through ENA imaging with IBEX have offered a tantalising glimpse of what may be achieved by global imaging. In particular, the ENA imaging provided estimates of exospheric densities in the vicinity of the magnetopause, a key parameter needed to assess the likelihood of success for both ENA and soft X-ray imaging.

The relatively recent discovery of SWCX X-ray emission provides a new path forward in the study of magnetospheric physics. As discussed by Robertson et al. (2006), electron charge exchange from a neutral exospheric atom to a heavy, high charge state solar wind ion leads to the emission of a photon at extreme ultraviolet or soft X-ray wavelengths. Such emission has been detected in the vicinity of the Earth's magnetosphere by XMM-Newton (e. g. Snowden et al. 2004, 2009; Carter et al. 2011) and Suzaku (Fujimoto et al. 2007, Ezoe et al. 2010); simulations have been performed demonstrating that SWCX emission is sufficiently bright to image the cusps and magnetosheath and to determine the location of boundaries such as the magnetopause and bow shock (Robertson et al. 2003a,b, 2006). As such, SWCX imaging provides a novel method by which we can in practice 'see' the magnetosphere for the first time and open up a new window to the study of solar wind – magnetosphere physics.

Below we discuss some of the outstanding key questions that the AXIOM mission will address. They are summarized at the end in Table 1, which shows the AXIOM Science Traceability Matrix, where science questions are matched to AXIOM observing strategy and objectives, and instrument requirements, with in situ measurements detailed in Table 2.

## 2.2 Magnetopause physics

**How do upstream conditions control the thickness of the magnetosheath and the size and shape of the magnetopause?**

The size and shape of the magnetopause and the thickness of the magnetosheath are key parameters defining the solar wind – magnetosphere interaction. Comprehensive knowledge of the global shape and position of the magnetopause and associated boundary layers as a function of the prevailing exterior conditions is still lacking. This is partly because in situ spacecraft generally only observe the magnetopause in transition from one state to another, causing statistical models for the shape and position of the magnetopause to exhibit large error bars (e.g. Fairfield 1971, Farris et al. 1991, Petrinec et al. 1991, Roelof and Sibeck 1994). Although data from the four spacecraft Cluster mission have provided accurate measurements of the instantaneous position and velocity of the magnetopause (Dunlop et al. 2008), these are values for a magnetopause in transition, not one in a stable position. Thus it is not simple to relate the measurements to the prevailing solar wind conditions, and the observations have not been able to lead to substantial improvements of the models for magnetopause or bow shock location as a function of solar wind conditions. Therefore the two key questions remain: 1) How do the position and shape of the global magnetopause boundary depend on solar wind conditions? 2) How does the thickness of the magnetosheath change with solar wind conditions? Global imaging is required to fully understand magnetopause morphology as a function of prevailing solar wind conditions. X-ray imaging of the dayside outer magnetospheric boundaries offers a way to obtain this information.

Because we seek to determine the size and shape of the dayside magnetosphere and magnetopause locations vary by several $R_E$, an X-ray imager with a wide field of view (FOV, ~10° x 10°, equivalent to ~10 $R_E$ x 10 $R_E$ at a distance of 50 $R_E$) located at a vantage point significantly outside the magnetosphere is required. The imager must be capable of distinguishing between SWCX emissions on lines of sight that pass through the magnetosheath and those that pass through the magnetosphere. The distance from the Earth to the subsolar magnetosphere varies approximately as the sixth root of the solar wind pressure, such that a 10% increase of the latter leads to a 2% reduction of the former – equivalent to a fraction of $R_E$. This implies that an imager spatial resolution capability of 0.1 $R_E$ is required (for the nose of the magnetosphere). The occurrence rate of solar wind discontinuities is thought to be ~3.6 /hr (Mariani et al. 1973), thus intervals of stable solar wind lasting ~10 – 15 min between discontinuities should be identified, and during this time the location of the magnetopause mapped. Given known limitations in propagating solar wind features from the Sun – Earth L1 point to the magnetopause, and given that there is no guarantee that in situ observations at L1 will be available in the AXIOM operations timeframe, it is therefore crucial that the mission baseline payload includes in situ solar wind plasma and magnetic field measurements. The payload should include a magnetometer to determine the strength and direction of the prevailing solar wind magnetic field with a high cadence to establish the orientation of solar wind boundaries.

**How does the location of the magnetopause change in response to prolonged periods of subsolar reconnection?**

Magnetic reconnection at the dayside magnetopause controls global magnetospheric structure and dynamics. The shape and location of the magnetopause boundary itself may be dramatically altered as a result of enhanced dayside reconnection and the generation of 'Region 1' currents. Southward turnings of the interplanetary magnetic field (IMF) enhance reconnection. Region 1 currents flow along the boundaries separating reconnected and closed magnetic flux. Consistent with expectations that reconnection removes magnetic field flux from the dayside magnetosphere, the Region 1 currents depress dayside magnetic field strengths and allow the magnetopause to move inward even in the absence of solar wind dynamic pressure variations. Precise quantification of this effect is crucial to constraining theories of solar wind – magnetosphere coupling, and understanding the transfer of solar wind energy into the magnetosphere on a global scale. The measurement objectives here are similar to those described above. The magnetic flux removed from the dayside magnetosphere accumulates within the magnetotail, where it is subsequently released during the course of a substorm, either in response to external solar wind triggers or internal instabilities. Substorms recur on the order of once every three hours, while significant dayside flux erosion events transpire over periods ranging from 30 min to 1 hour. A global imager with a cadence on the order of several minutes (as required by the previous scientific objective) will suffice to track the erosion events that occur following southward turnings of the IMF.

**Under what conditions do 'transient' boundary layers, such as the plasma depletion layer, arise?**

In situ measurements reveal that boundary layers with plasma properties intermediate between those of the magnetosheath and magnetosphere can frequently (but not always) be found bounding the magnetopause current layer. Within the magnetosheath itself, a plasma depletion layer (PDL) often forms as the IMF piles up ahead of the magnetopause (e.g. Wang et al. 2004). Within this layer, the plasma is 'squeezed out' along the magnetic field, resulting in densities depressed from those in the upstream magnetosheath and, to maintain pressure balance, enhanced magnetic field strengths. Sporadic observations of the PDL indicate that its thickness and the degree of plasma depletion within it are highly variable, most likely depending on the prevailing solar wind and magnetopause conditions. Since conditions within the PDL are those actually applied to the magnetosphere, and may differ greatly from those in the magnetosheath proper, a clear understanding of the PDL is essential to determining whether or not reconnection can occur at the magnetopause, and if it does occur, what form it takes. For example, the presence of a depletion layer enhances Alfvén velocities, reducing Mach numbers, and increasing the likelihood of steady (as opposed to pulsed) reconnection.

Magnetopause reconnection accelerates particles along field lines, forming energy-dispersed layers of particles with magnetosheath origin just earthward from the magnetopause, such as the high latitude boundary layer (HLBL). At more equatorial latitudes, there is also a low latitude boundary layer (LLBL), which has been variously attributed to reconnection, diffusion, the non-linear Kelvin-Helmholtz instability and other mechanisms governed by solar wind conditions.

Although we are aware of their existence, the instantaneous global extent and thickness of these boundary layers (extent along the magnetopause, thickness as a function of solar wind parameters, etc.) has not been determined due to the lack of global coverage of their structure. Again, X-ray imaging offers a methodology to address these issues, with measurement objectives similar to those already described. By studying the layers, we can learn much about the occurrence patterns, extent, and therefore significance of microphysical processes governing the solar wind – magnetosphere interaction.

## 2.3 Cusp physics

Knowledge of the location, size and shape of the cusps is crucial for understanding how solar wind plasma is transferred into the magnetosphere; in situ measurements are limited in the extent to which they can resolve spatio-temporal ambiguities. The high altitude cusps have not previously been imaged, so the instantaneous global extent of the cusps in latitude and longitude, extending from the ionosphere to the magnetopause, has not been observed, resulting in severe ambiguities (Fritz and Zong 2005). A primary objective of AXIOM will be to image the location and structure of the cusps for a variety of solar wind speeds and densities and IMF orientations, in order to understand the shape of the dayside magnetosphere under different coupling conditions, free of spatio-temporal ambiguities. By viewing the cusps from the side it will be possible to probe their latitudinal and altitude structure, while viewing from the front will allow the local time position to be observed. Key questions relate to the width of the cusps both in latitude and local time.

Spacecraft flying latitudinally through the cusps have revealed that their structure can be (apparently) filamentary in latitude, leading to the suggestion that there can be multiple cusps (e.g. Zong et al. 2008). It is possible that this spatial structure can exist, or that the appearance of multiple cusps is caused by movements of the cusp over the spacecraft as the open flux content of the magnetosphere waxes and wanes. Imaging will reveal whether this structure is real or apparent. As for the magnetopause, a wide FOV X-ray imager, observing from a point well outside the magnetosphere, is required, capable of detecting the SWCX emission from the cusps. The cusps are expected to be the brightest sources of SWCX emission (Robertson et al. 2006), owing to their high densities and relative proximity to Earth, where the exospheric density is higher, and a higher time cadence of measurement is envisaged.

**How do the cusps move in response to changes in the solar wind?**

The latitudinal location of the cusp depends on the level of interconnection of the Earth's dipole with the IMF, i.e. on the amount of 'open' magnetic flux in the magnetosphere (Milan et al. 2003). The cusps, being associated with the boundary between open and closed field lines regions at the dayside, should move to lower and higher latitudes as the open field line region expands and contracts. The different viewing conditions during the AXIOM orbit will enable observing how the E-W component of the solar wind magnetic field controls cusp location and how solar wind coupling alters cusp latitude.

**How does the cusp density depend on magnetospheric coupling?**

The cusp density provides information about the extent to which solar wind plasma can access the magnetosphere. Such large scale properties are difficult to determine un-ambiguously from in situ measurements, while imaging can show changes in the brightness and thus density of the cusps for different solar wind conditions (like speed, or density, or orientation of the IMF) and dipole tilt angles. At times other than the equinoxes, the cusps are not symmetrically presented to the incoming solar wind; it is possible that this will change the relative penetration of material into the two cusps. Such changes will reveal themselves in the SWCX emission from the cusp regions.

Magnetic reconnection can be pulsed in nature, either as a consequence of changes in the solar wind or due to the natural variability of the reconnection process itself. Subsolar reconnection can appear to occur as quasi-periodic bursts, known as flux transfer events or FTEs (Russell et al. 1978), with a canonical repetition rate close to 8 min (Rijnbeek et al. 1984). The newly-reconnected flux tubes associated with each FTE should travel across the magnetopause to high latitudes at a speed of ~100 km s$^{-1}$ and can be imaged as corresponding inward jumps in magnetopause location, and enhancements in soft X-ray emissions caused by the particles entering the cusps. Such observations can resolve questions about the local time extent of FTEs, and by inference the time-dependent nature of reconnection, which is critical for understanding the overall solar wind – magnetosphere coupling process.

## 2.4 Shock physics

**What controls where the bow shock forms upstream of a planetary magnetosphere?**

One of the most basic and fundamental issues concerning the solar wind – magnetosphere interaction is the location and the extent of the bow shock. Evidently, the location of the bow shock relative to the planet depends on the solar wind conditions; typically it is observed a few $R_E$ upstream of the Earth's magnetopause, but its actual location, and shape, can vary considerably with upstream conditions.

Theory predicts that the location of the bow shock depends on that of the magnetopause. The magnetopause location depends on the pressure balance between the solar wind and the magnetospheric magnetic field. Cases of simultaneous bow shock and magnetopause motion have been reported, such as the coherent excursion observed with simultaneous in situ measurements by Cluster and Double Star (Cai et al. 2009); however, these are fortuitous occurrences, and the global extent, together with the steady-state locations of these boundaries as a function of varying solar wind conditions, remain very uncertain. Consequently it is very difficult to distinguish between different theories of bow shock behaviour (Fairfield et al. 2001). This is a crucial issue, because in situ studies of shock processes (e.g. particle acceleration upstream of the shock) require a model to determine the location of the shock.

Imaging of the shock can be achieved using SWCX X-rays, since emission along lines of sight that pass through the denser magnetosheath will be greater than that entirely in the solar wind. To observe the bow shock, magnetosheath and magnetopause together requires an observation point outside the magnetosphere, approximately perpendicular to the Sun – Earth line. An artist's impression of this observing scenario is offered by the AXIOM logo in Fig. 3 (note that the Earth is not to scale and the cusps are missing).

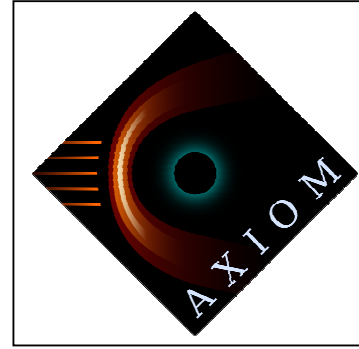

A wide FOV imager, from a vantage point well outside the magnetosphere, is again required, with a spatial resolution capability of ~0.25 $R_E$. As for the magnetopause (sec. 2.2), a steady state timescale of ~10 – 15 min is expected, during which shock and magnetopause reach their equilibrium position.

**How does the steady state thickness of a collisionless shock depend on the upstream conditions?**

Theory predicts that the thickness and structure of collisionless shocks depend strongly on upstream conditions. In particular, the angle between the magnetic field and the shock normal is crucial. Roughly speaking, quasi-perpendicular shocks are thin, with shock structure of the order of characteristic ion-scales. Quasi-parallel shock transitions are much more complex and extended (Lucek et al. 2005, 2008).

**Fig. 3** AXIOM logo: Artist's impression of SWCX X-ray emission from the magnetosheath (the Sun is to the left; note that the Earth is not to scale and the cusps are missing)

The configuration over most portions of the shock is expected to be oblique (i.e. with angles between the IMF and the shock normal between 0° and 90°), and exhibit complex structures. Several theories explaining quasi-parallel shock physics have yet to be tested properly due to difficulties in studying the extended structure of these shocks with in situ measurements. Global imaging has the advantage that the boundary can be identified as the location where line of sight emissions increase from low levels in the solar wind to high levels in the magnetosheath. The required spatial resolution is the most stringent condition, being at least 0.1 $R_E$. By imaging the shock directly, for a variety of solar wind speeds and densities, it will be possible to determine its thickness as a function of upstream flow conditions.

## 2.5 Interaction of a CME with the magnetosphere

Coronal mass ejections (CMEs) are often characterised by low proton temperatures, enhanced alpha/proton ratios, strong magnetic fields, smooth magnetic field rotations and unusual heavy ion ionisation states (e.g. Aguilar-Rodriguez et al. 2006, and references therein). CMEs play a key role in causing strong geomagnetic storms at Earth. There is significant density variation inside CMEs, which has been detected remotely through soft X-ray emissions measured by XMM-Newton, as shown in Fig. 4 (Carter et al. 2010). CMEs also drive shocks which correspond to a step change in the solar wind dynamic pressure. Such shocks cause dramatic rapid compression of the magnetosphere leading to prompt trapping of solar energetic particles in the inner magnetosphere (Hudson et al. 2004).

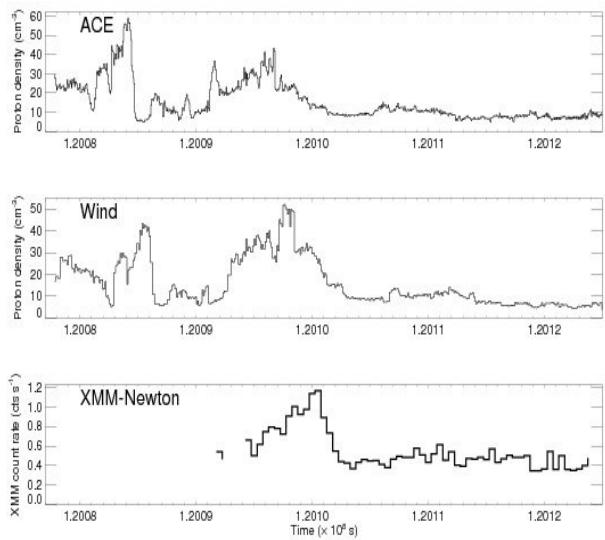

**Fig. 4** XMM-Newton detection of a CME (Carter et al. 2010). Using SWCX X-ray emission, XMM-Newton detected the enhancement in solar wind density that was observed in situ by the ACE and Wind spacecraft.

A CME, moving from upstream and then interacting with the bow shock and magnetosheath over a period of ~ 1 day, provides a perfect opportunity to perform X-ray spectroscopic analysis since the charge states of the plasma are expected to be different in different regions of the CME plasma. This gives remote information about CMEs structure, which is crucial for understanding their formation and evolution. Imaging at a cadence of 5 – 15 min, over the whole duration of a CME, will allow the entire structure of the CME within the FOV to be determined. As the composition of the CME changes, the relative brightness of different X-ray emission lines will change.

## 2.6 Additional secondary science

For an orbit not far out from the Earth's equatorial plane, such as planned for AXIOM (see sec. 3), there will be times when the view of the nose of the magnetosphere is contaminated by the bright Earth or the Sun. For at least some of these periods the X-ray imager can be positioned in such a way as to observe the flanks of the magnetosheath avoiding the Earth. Another tantalising possibility is that of imaging the lobe/magnetosheath boundary, i.e. the tail magnetopause; this would give a measure of how wide the tail is in the N – S direction, and how populated, for different solar wind conditions. It will be feasible to attempt this for the near-Earth magnetotail since the growth phase of substorms, dayside erosion, and consequently magnetotail flaring in response to the addition of magnetic flux removed from the dayside, is 30 min to 1 hour, i.e. longer than the required image accumulation times.

When observing close to the Earth is impossible, time can also be used for 'opportunistic' observations, such as those of comets that may fly by at 1 – 2 AU. Comets, with their extended neutral comae, are ideal targets for SWCX, and act as probes of solar wind conditions at different distances in proximity of the Sun (Bodewits et al. 2007). The rest of the time will be spent calibrating the imager performance, using 'standard candles' such as supernova remnants.

**Table 1 AXIOM Science Traceability Matrix**

| Science Question | AXIOM Strategy and Objectives | Imaging Requirements | In situ Requirem.s (see Table 2) |
|---|---|---|---|
| **1.1 Magnetopause physics:** *How do upstream conditions control magnetopause position and shape and magnetosheath thickness?* | Identify intervals of quasi-steady solar wind conditions lasting ~10 - 15 min and image the location of the magnetopause during each steady interval. Establish which, if any, of current magnetopause models most accurately explain observations. | Wide FOV (10 $R_E$ scale) SWCX imaging from outside the magnetosphere. Spatial resolution of 0.1 $R_E$ at a cadence of 15 min. | Solar wind composition, solar wind plasma, solar wind magnetic field |
| **1.2 Magnetopause physics:** *How does the location of the magnetopause change in response to prolonged periods of subsolar reconnection?* | Image the location and thickness of the magnetopause in intervals following southward turning of the solar wind magnetic field triggering subsolar reconnection. Image the changing location of the magnetopause and thus determine how the magnetopause moves in response to southward turning. | Requirements similar to Q1.1, but with time cadence of the order of a few minutes. | Solar wind magnetic field, solar wind plasma, solar wind composition |
| **1.3 Magnetopause physics:** *Under what conditions do transient boundary layers, such as the plasma depletion layer (PDL), arise?* | Image the occurrence of the PDL and other transient boundary layers in the course of magnetopause observations, and parameterise occurrence as a function of upstream solar wind conditions. Use imaging to determine the spatial extent and thus establish its role modulating solar wind – magnetosphere coupling. | Requirements similar to those described above in Q1.1. | Solar wind plasma, solar wind magnetic field, solar wind composition |
| **2.1 Cusp physics:** *Cusp morphology - what are the size and shape of the cusps?* | Image the cusps and their internal structure as a function of upstream solar wind conditions. Establish the homogeneity of the cusps and determine whether filamentary structure observed by spacecraft is spatial or temporal in nature. | Wide FOV (10 $R_E$ scale) SWCX imaging from outside the magnetosphere. Spatial resolution of 0.1 $R_E$ at a cadence of 1 min. | Solar wind plasma, solar wind magnetic field, solar wind composition |
| **2.2 Cusp physics:** *How do the cusps move in response to changes in the solar wind?* | Image the location of the cusps and track their motion in response to changes in the solar wind magnetic field orientation and the solar wind dynamic pressure. | Requirements similar to those described in Q2.1. Imaging at angles oblique (i.e. less than 90°) to the Sun – Earth line will enable study of the E – W motion of the cusp. | Solar wind magnetic field, solar wind plasma, solar wind composition |
| **2.3 Cusp physics:** *How does the cusp density depend on magnetospheric coupling?* | Image the density of the cusp as a function of season to determine the asymmetries in plasma entry through the Northern and Southern cusps. Image the response of the cusp to changes in the solar wind magnetic field orientation and the onset/cessation of magnetic reconnection at the dayside magnetopause. | Requirements similar to those described in Q2.1. Images of the cusp to be acquired over more than 1 year to determine its structure as a function of dipole tilt angle. | Solar wind plasma, solar wind magnetic field, solar wind composition |
| **3.1 Shock Physics:** *What is the location of the bow shock for given solar wind conditions?* | Identify intervals of quasi-steady solar wind conditions lasting ~10 - 15 min (based on typical convection time of solar wind over the dayside magnetosphere and occurrence rate of solar wind discontinuities). Image location of bow shock and magnetopause during each steady interval. Build up catalogue of observations. Establish which, if any, of current shock models most accurately explains observations. | Wide FOV (10 $R_E$ scale) SWCX imaging from outside the magnetosphere. Simultaneous imaging of the location of the dayside subsolar magnetopause and bow shock. Spatial resolution of 0.25 $R_E$ at a cadence of 15 min. | Solar wind composition, solar wind plasma, solar wind magnetic field |
| **3.2 Shock Physics:** *What is the steady-state thickness of the quasi-parallel collisionless shock as a function of upstream conditions?* | Identify intervals of quasi-steady solar wind conditions lasting ~10 - 15 min (see above). Image the thickness of the bow shock transition. Establish the typical large scale thickness of the quasi-parallel shock as a function of magnetic field orientation, plasma β (ratio of plasma pressure to magnetic pressure) and solar wind Mach number. | Image the bow shock with a spatial resolution of 0.1 $R_E$ at a timescale of 15 min. Image should encompass the region of the bow shock where the magnetic field is quasi-parallel to the shock normal. | Solar wind magnetic field, solar wind plasma, solar wind composition |
| **4 CMEs:** *How do Coronal Mass Ejections interact with the magnetosphere?* | Identify CME signatures in the in situ plasma data. Use spectroscopy to determine the spatial extent of different regions of the CME based on composition. Image the changing location of the bow shock and magnetopause to determine magnetospheric compression in response to a CME. | Spectroscopic X-ray imaging of the interaction of CMEs and the magnetosphere, capable to distinguish between different expected types of solar wind composition in CMEs. Time cadence of 15 min and spatial | Solar wind composition, solar wind plasma, solar wind magnetic field |

| | | resolution of 0.5 $R_E$, and should capture the complete passage of the CME through the FOV (CME duration at 1 AU ~ 1 day). | |

**Table 2 AXIOM Science Traceability Matrix (in situ measurements)**

| Measur. Type | Rationale | Measurement Requirements | Parameters | Tempor. Cadence |
|---|---|---|---|---|
| Bulk solar wind | Required to determine the solar wind input controlling shock, magnetopause and cusp location (plasma β and solar wind Mach number) | 3D distribution / moments (density, velocity, temperature) of H+ and He++, sufficient to resolve solar wind structure | < 20 keV/q in energy (where q is the particle charge) | 3 s |
| Solar wind compos. | Required to properly calculate the nature of the charge exchange output | Distribution of minor ion species on a timescale comparable with X-ray imaging | Minor ions important for SWCX: C5+, C6+, N6+, N7+, O7+, O8+, Fe17+, Fe18+, Mg11+, Mg12+ | 300 s |
| Magnetic field | Required to determine the magnetic field geometry at bow shock, magnetopause and in the cusps. Occurrence/orientation of solar wind discontinuities | Orientation and strength of solar wind magnetic field | 0.25 nT accuracy, 10 pT precision | 16 vector/s |

# 3 AXIOM mission profile

AXIOM will carry a payload specifically targeted to achieving the scientific objectives set out in sec. 2, comprising an X-ray imaging and spectroscopy instrument, complemented by a compact plasma package and a magnetometer. This allows in situ measurements of the solar wind to be carried out simultaneously with the X-ray observations. Depending on the location of AXIOM and assuming average solar wind speeds of 400 - 600 km s$^{-1}$, lags from the time of solar wind observation at AXIOM (60 $R_E$ from Earth) to the SWCX emission response at locations some 10 $R_E$ upstream from the dayside magnetosphere will range from 0 min (when AXIOM is located at positions near the Earth's orbit around the Sun) to 10 - 15 min (when AXIOM is located directly upstream from the Earth's magnetosphere). In practice, the plasma measurements can be matched directly to the X-ray emissions.

The baseline payload does not require a large spacecraft, thus a Vega launcher has been selected, which also minimises the cost of the mission. Preliminary mission analysis shows that a Vega launcher along with a separate propulsion module will take the required payload mass to the final operational orbit. The LISA Pathfinder Propulsion Module (PRM) is considered a good candidate for re-use for AXIOM.

In order to obtain a global view of the Earth's magnetosphere we require a vantage point far out from the planet, at a distance of the order of tens of $R_E$. Several orbit possibilities, both equatorial (circular) and polar (elliptical), have been examined. Equatorial orbits provide the 'classical' view from the side of the magnetosphere for a large fraction of the time, but suffer from having the (far too X-ray bright) Earth in the FOV for part of it. Polar elliptical orbits are much better at avoiding viewing the Earth, but imply regular passages through the Earth's radiation belts (unless perigee is above ~ 6 $R_E$), which impose more severe radiation shielding requirements. In both configurations, we require that the Sun angle be above 45° from the edge of the FOV, to help avoid stray-light affecting the X-ray optic.

A trade-off has been carried out in order to establish the best orbit for the mission. Various polar and equatorial orbits, as well as one with the spacecraft in a Lissajous orbit around the Earth – Moon L1 point, have been considered. The severity of the radiation dose, as well as approximate observing efficiencies, derived for observing the nose of the magnetosphere directly, have been determined for each orbit; the mass that can be taken to orbit by the Vega launcher and PRM combination has also been calculated for each case, and all the results have been entered in the trade-off. Equatorial orbits with the largest radii and the L1 orbit score highest; the L1 Lissajous orbit was eventually selected because it allows the largest mass (451 kg) to be taken to orbit. However other lunar, or near lunar, orbits are thought to be a possibility. It is also feasible to increase the observing efficiency from such an orbit by pointing at other parts of the magnetosheath, away from the nose, during bright-Earth phases.

Fig. 5 illustrates the AXIOM Lissajous orbit geometry, the viewing angle for different positions in the orbit, the operational phases for primary and secondary science and the observing constraints.

The Vega launcher will take the science spacecraft and the PRM into a highly elliptical orbit from which the propulsion module would perform a series of apogee raising manoeuvres to approximately 1.4 million km. At this point, Weak Stability Boundary effects due to solar gravity perturbations start playing a role, and will raise the perigee to approximately lunar orbit radius. This low speed approach to the Moon allows a 'weak capture' which can then be manipulated to achieve a Lissajous orbit around L1.

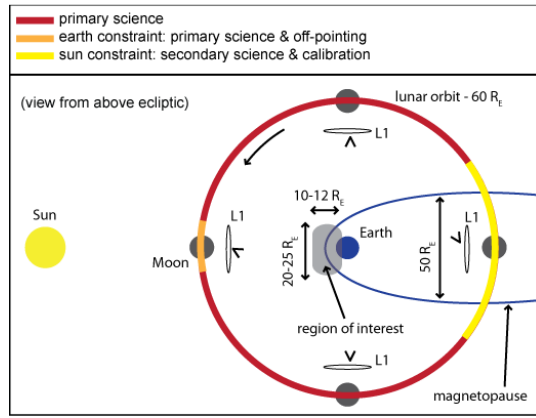

**Fig. 5** AXIOM operational orbit around the Earth-Moon L1 point (not to scale): the narrow ellipses near the Moon represent the Lissajous orbit; marks such as ^ indicate the viewing angles.

The full transfer trajectory would take between 5 and 7 months. Once in the Lissajous orbit, periodic station-keeping is needed, approximately every 7 days; this is not a serious drawback (e.g. the NASA ARTEMIS spacecraft currently operates in such an orbit without reported problems). The period of the orbit is about 14 days, with the larger L1 orbit around the Earth being 28 days (see Fig. 5).

A single 15 m X-band ground station is envisaged for AXIOM telemetry, tracking and command, which will allow for an average of 8 hr/day downlink time, with some variation with the season. The nominal lifetime of the mission to achieve its science goals is 2 years, with an extended mission of another 2 years highly recommended for science purposes.

# 4 Proposed model payload

AXIOM's payload comprises an X-ray imaging and spectroscopy instrument, together with a compact plasma package and a magnetometer, so that in situ measurements of the solar wind can be carried out simultaneously with the X-ray observations. The main components of the X-ray imager (WFI – Wide Field Imager), which will observe the Earth's magnetosphere continuously for most of the time, are light-weight optic with a large FOV and large area focal plane detectors. The plasma package incorporates a proton-alpha sensor (PAS) and an ion composition analyser (ICA); these have a 360° FOV so that they can monitor the solar wind parameters independently of the WFI pointing direction. A magnetometer (MAG) completes the payload, to obtain continuous measurements of the strength and orientation of the solar wind magnetic field.

## 4.1 X-ray Wide Field Imager (WFI)

### 4.1.1 Measurement technique

The primary science target of AXIOM is an extended structure surrounding the Earth emitting X-rays at characteristic line energies in the range 0.1 to 2.5 keV via the charge exchange process. From the Earth – Moon L1 position this structure, encompassing the bow shock, magnetopause and magnetospheric cusps, has a scale size of several degrees. The key requirements for the X-ray WFI are:

1) Sensitivity to X-rays in the energy range 0.1 to 2.5 keV (primary science target), and to higher energies for calibration purposes,

2) An energy resolution of < 65 eV (FWHM) at 0.6 keV sufficient to isolate the major X-ray emission lines (primarily from OVII and OVIII in the energy band 0.5 to 0.7 keV),

3) Imaging capability with an angular resolution of ~7 arcmin (equivalent to a scale size of 0.1 $R_E$ at a distance of 50 $R_E$),

4) A wide FOV (baseline of 10º x 15º),

5) Time resolution of ~1 minute.

These requirements can be met by a telescope which couples a microchannel plate (MCP) optic array with a detector plane employing X-ray sensitive CCDs.

### 4.1.2 Instrument conceptual design and key characteristics

Table 3 summarises the key characteristics of the X-ray WFI. In practice, the wide FOV of the X-ray imager can only be achieved within the mass budget by the use of MCP optics. This technology is used for the optic of the Mercury Imaging X-ray Spectrometer (MIXS) on ESA's BepiColombo mission due for launch in 2014. Leicester University is the PI institute for MIXS and a world leader in developing this technology for astrophysics applications in association with the industrial manufacturer of the glass plates, Photonis France S.A.S[1].

The basic focusing geometry of an MCP optic is shown in Fig. 6 (left panel). The complete optic is manufactured from individual glass plates, typically 4 cm x 4 cm and 1 mm thickness. The plates contain a square-packed array of pores (typically 20 µm in width with wall thickness of 6 µm) and the X-ray focusing is achieved by grazing incidence reflection off the sides of the pores. Each plate is heat-slumped over a former to the desired radius of curvature. The segments are then placed in a frame structure to give mechanical strength (Fig. 6, right panel).

**Table 3 X-ray WFI key characteristics**

| Attribute | Value | Notes |
|---|---|---|
| Optic FOV | 10° x 15° | |
| Optic angular resolution | 2 arcmin FWHM | |
| Optic focal length | 70 cm | |
| Optic effective area at 0.6 keV | 58 cm$^2$ | Assuming Iridium coating |
| Total instrument effective area at 0.6 keV | 37 cm$^2$ | Assuming CCD QE ~ 0.8 and 1600Å Polyimide + 800Å Al UV filter |
| Detector energy resolution at 0.6 keV | 65 eV FWHM | Assuming 6 e$^-$ readout noise |
| Detector pixel size (effective) | 60 μm | Device/Mode dependent |
| Time resolution | 31.1 sec | 30 s frame time + 1.1 s readout time |

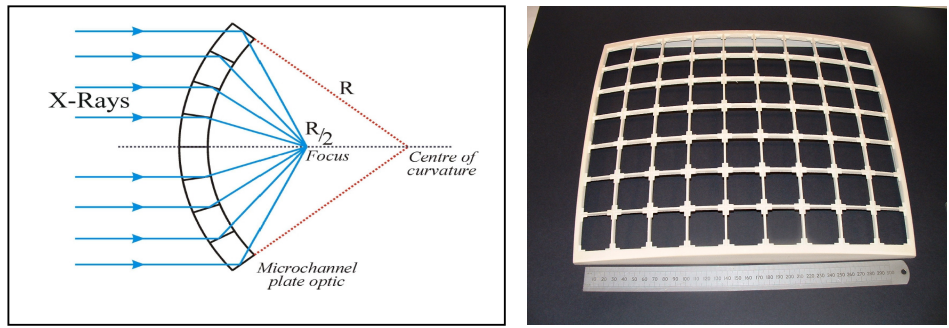

**Fig. 6** Left: Basic focusing geometry of MCP optic. Right: Example frame constructed at Leicester University for holding individual MCP plates (scale is 30 cm across).

The channel walls of the MCPs can be metal-plated with, for example, Iridium or Platinum to improve the X-ray reflectivity compared with bare glass within the key energy range 0.5 – 0.7 keV. The mass of the entire optic plane for this instrument is < 1 kg. Individual plates of this configuration have already been measured in the laboratory at Leicester University (G. Fraser, private communication) to have a point spread function (PSF) of ~2 arcmin (FWHM). This is well within the performance required for the primary science goals of the mission. Fig. 7 shows data from one such measurement. X-rays which undergo a double reflection from orthogonal channel walls are brought to the prime focus. The cross arms of the PSF are due to rays which only have a single reflection off one of the channel walls.

The baseline instrument FOV is a rectangle of 10º x 15º which provides a good compromise for encompassing the primary region of scientific interest whilst excluding the bright Earth (sec. 3).

---

[1] http://www.photonis.com

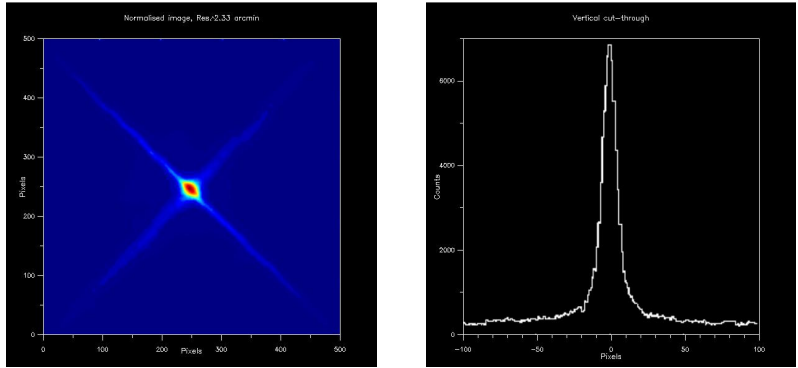

**Fig. 7** Laboratory-measured MCP optic's Point Spread Function. Plate scale is ~ 5 pixels/arcmin. FWHM is ~ 2 arcmin. Data taken with the 27m beamline at Leicester University.

Fig. 8 shows one result from simulations of AXIOM's viewing of the nose of the magnetosheath; such simulations are invaluable in determining viewing constraints, and establishing the observing efficiencies for different orbits and at different orbital phases (see http://www.star.le.ac.uk/~amr30/AXIOM/ for further examples and details of the simulations).

The instrument sensitivity is a function of the effective area of the optic, which scales with the focal length. As the instrument mass and cost also scales with the required FOV and the focal length, the design is a compromise between these factors and the scientific goals of the mission. Our baseline design has a focal length of 70 cm. With the specified FOV this leads to an optic size of about 24 cm x 36 cm or 6 x 9 = 54 of the 4 cm plates discussed previously.

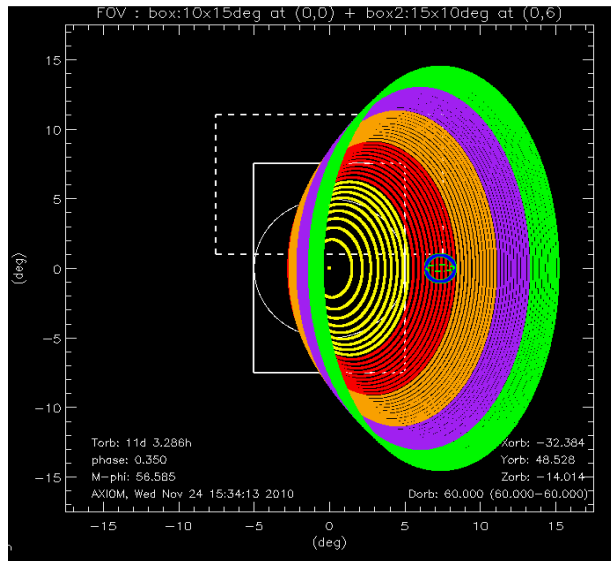

**Fig. 8** Visualisation produced with the AXIOM magnetopause/FOV simulator.

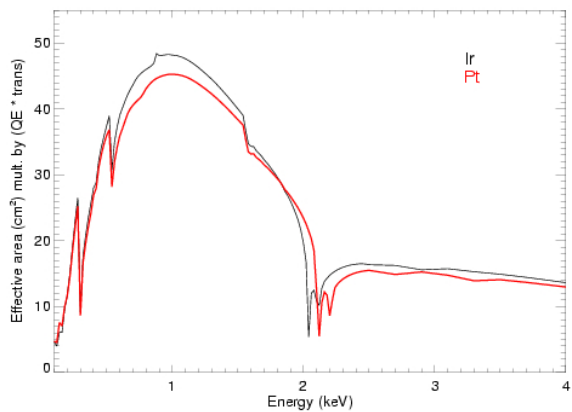

**Fig. 9** WFI predicted total effective area for both an Iridium and a Platinum coated optic.

The detector must be half the size of the optic to cover the FOV, hence the detector plane must be 12 cm x 18 cm in size given the optic geometry. The current design consists of a tile of wide-area X-ray sensitive CCDs, with the CCD231-84 back-illuminated CCD manufactured by e2v[2] as the baseline device; its image area of 6.14 cm x 6.14 cm leads to the detector requiring 6 of the devices in a 2 x 3 array.

The focal plane of the optic is on a hemisphere of half the radius of curvature of the optic array. Because the detector plane is constructed from large area CCDs, which are flat, the plane must deviate from the optimum focus in some areas even if the devices are angled with respect to each other. We have calculated via simulations that even if the whole detector plane were constructed to be flat, the angular resolution would only degrade to ~4 arcmin (FWHM) at the point furthest from the true focal plane. Fig. 9 shows the WFI combined effective area (optic effective area multiplied by the detector efficiency).

---

[2] http://www.e2v.com

For the detector efficiency we have used the quantum efficiency (QE) typical of back-illuminated CCDs, i.e. those of the EPIC-pn camera on XMM-Newton (Strüder et al. 2001), multiplied by a representative transmission curve of a UV filter of the same type as the EPIC Medium filter (1600 Å polyimide plus 800 Å aluminium), which is capable of blocking optical contamination from stars as bright as $m_v = 6 - 9$.

The imager will have a fixed filter deposited directly onto the CCDs. Optical loading degrades the energy resolution and alters the local charge background distribution across the CCD, thereby shifting the X-ray event energy. Measuring the charge distribution in the pixels surrounding each X-ray event can mitigate the latter effect, as it is done within the event processing of the EPIC-MOS cameras on XMM-Newton.

### *4.1.3 Performance assessment*

The performance of the WFI has been estimated using the simulations of the expected X-ray emissivity of the magnetosheath (and surrounding terrestrial volume) by Robertson et al. (2003a, 2006) for two states of the solar wind: a quiescent solar wind (density ~7 cm$^{-3}$, speed ~400 km s$^{-1}$) and a state representative of a storm (CME) which occurred in March 2001. The simulations predict peak emissivities of 8.8 and 160 keV cm$^{-2}$ s$^{-1}$ sr$^{-1}$ respectively (note the quiescent state model has no simulation of the cusp structure).

Fig. 10 (left panels) shows predicted WFI detector images derived from the SWCX emissivity maps folded through the detector response (assuming a viewing distance of 51 $R_E$ from the Earth) with an expected background component added. The integration times are 100 s and 1 ks for the storm simulation and 1 ks and 10 ks for the quiescent state simulation. The scale of these images is counts ks$^{-1}$ per 0.1° pixel. Each image has been smoothed to bring out detail. We also show (right panels) the significance of the SWCX emission in number of sigmas above the estimated background. The significance maps are binned on a 0.5° pixel scale. The background is a combination of the typical diffuse sky background, ~8 keV cm$^{-2}$ s$^{-1}$ sr$^{-1}$ at 1 keV (Lumb et al. 2002), and an estimated instrumental background of ~10$^{-4}$ cts s$^{-1}$ keV$^{-1}$ per square arcmin at 1 keV (Kuntz and Snowden 2008), both derived from XMM-Newton data. In addition we expect bright Galactic and extragalactic point sources within the FOV. These will easily be identified and removed as point sources because of the ~2 arcmin (FWHM) angular resolution of the optic and the fact that the vast majority will be in known locations via the ROSAT all-sky survey.

More simulated images can be found at http://www.star.le.ac.uk/~jac48/axiomsims/

The peak count rates from the SWCX emission are ~25 (storm) and ~1.2 (quiescent) cts ks$^{-1}$ within a 0.1° pixel, compared with an estimated background rate of ~7.5 cts ks$^{-1}$ in the same pixel. In the storm simulation the global structure of the magnetosheath is easily visible above the background in only 100 s. The position and extent of the cusps, for example, should be detectable to within a fraction of an $R_E$ on timescales of a few minutes during periods of high solar wind densities. During average solar wind conditions integration times of ~1 ks would be required.

The WFI has 25 times the grasp (defined as effective area times FOV area) at 0.6 keV compared with XMM-Newton. It therefore has enormous collecting power for diffuse flux. In 1 ks the predicted number of counts from the SWCX component within the FOV is ~160,000 and ~10,000 respectively from the storm and quiescent solar wind simulations (compared with ~110,000 from the background components).

Despite the low pixel-by-pixel significance of the quiescent solar wind image in Fig. 10, the 1 ks exposure is very valuable as it demonstrates the approximate timescale over which the global magnetospheric structure is visible against the background (with a signal-to-noise ratio of 30 for the total signal in the FOV). The 10 ks exposure during quiescent wind conditions will enable comparisons on a 2 – 3 hour timescale, for which spacecraft stability is not an issue (with the pointing being known to 15 arcmin accuracy, notwithstanding any post-attitude reconstruction – see sec. 5.1).

An issue that will need addressing (and is under study, but whose solution is beyond the scope of this paper) stems from the fact the the signal, for a given line of sight, is the integrated emission of an extended, 3D tenuous medium. Retrieval of physical parameters, such as ion density, requires image inversion (assuming the exospheric density is known), which will need validating, e.g. by in situ measurements, as shown by Vallat et al (2004).

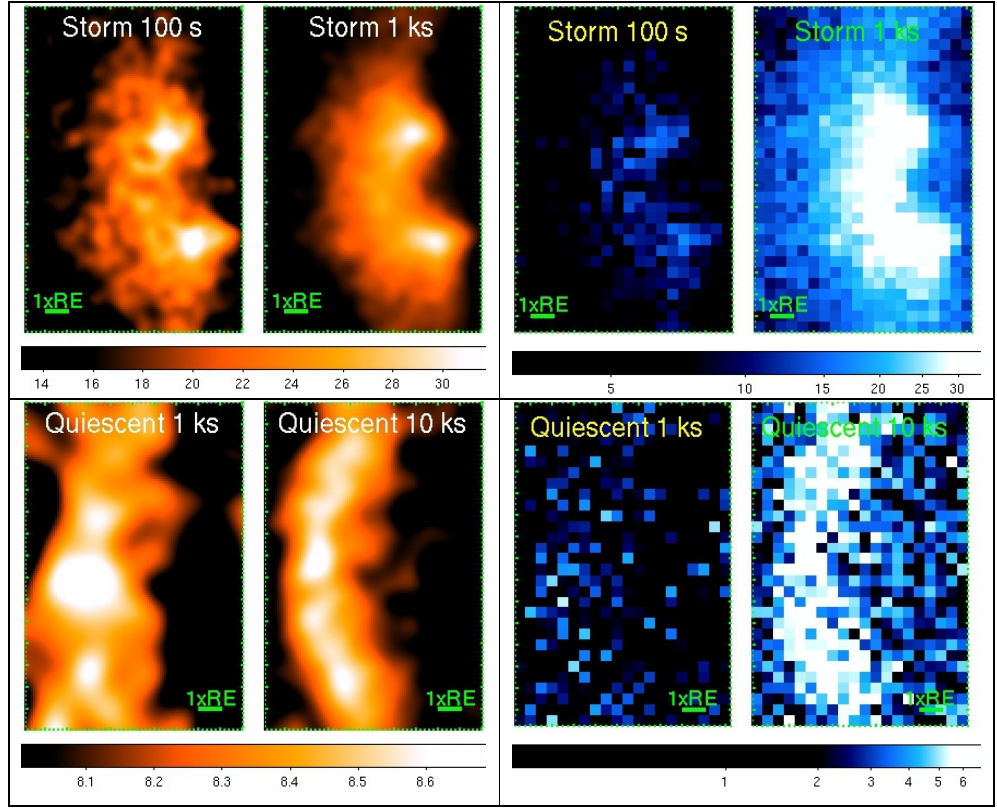

**Fig. 10** Simulated WFI images (left panels) and significance maps (right panels) for storm (upper panels) and quiescent (lower panels) solar wind conditions. The solar wind flows in from the left and the magnetosheath and cusps, bounded by the bow shock (left) and magnetopause (right), emit brightly via SWCX.

The WFI will also be able to accumulate high precision SWCX spectra, resolve the lines from the major ions and derive solar wind abundances for any selected part of the images (the CCD camera measures the energy, as well as the position, of every X-ray photon detected). Fig. 11 shows background-subtracted simulated spectra extracted from the whole FOV for 1 ks exposures assuming (left panel) a spectrum characteristic of a quiescent solar wind (Bodewits et al. 2007) and (right panel) a spectrum characteristic of a CME (Carter et al. 2010). The normalisations of the SWCX spectral model are derived from the above simulations. The charge state distribution varies considerably with solar wind type, as can be seen for example in the change between the OVII/OVIII ratios in Fig. 11.

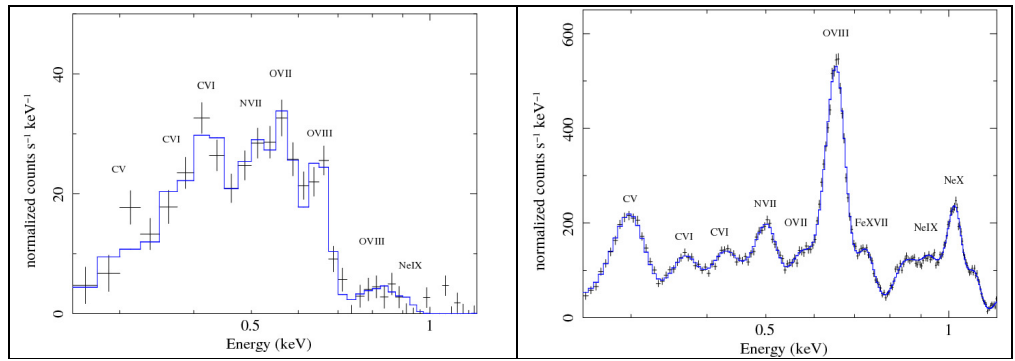

**Fig. 11** Simulated background-subtracted SWCX spectra (1 ks exposure, total WFI FOV) for (left panel) a quiescent solar wind and (right panel) a CME.

### 4.1.4 Interfaces, pointing and alignment requirements

Cooling of the WFI focal plane will be achieved with a passive radiator, which will need to dissipate around 13.3 W of heat load and achieve a baseline focal plane temperature in the range between -90° and -100°C. A heater system will maintain thermal control.

The WFI pointing stability requirement is better than 1 angular resolution element per frame integration time, which is approximately 2 arcsec per second. The instrument will need to be pointed with an accuracy of around 15 arcmin at the theoretical position of the bow shock. During favourable observing conditions (sec. 3) the long axis of the FOV needs to be aligned at an angle of 90° from the line between the Earth and the bow shock to an accuracy of a few degrees.

### 4.1.5 Operating modes

The CCDs will operate in full frame mode with a frame integration time of no more than 30 s.

The baseline CCD231-84 device has 4096 x 4412 15 μm square pixels with a readout noise of 6 e$^-$ at 2 MHz. This pixel size oversamples the optic's PSF by a considerable margin as the plate scale is 1 arcmin = 200 μm. The optimum pixel size is around 60 μm; this would allow for information on the local background charge distribution surrounding each X-ray event to be analysed as an aid to calibration of the event energy. The readout could therefore be binned into 4x4 pixels (effective pixel size = 60 μm) and operated in dual node readout mode. At 2 MHz the readout time would be about 1.1 s, giving an out-of-time (OTT) event rate of 3.7% for a 30 s frame integration time (cf EPIC-pn with an OTT rate of 6.3% in full frame mode).

The count rate for 60 μm pixels will be less than $10^{-3}$ cts pixel$^{-1}$ s$^{-1}$ for the SWCX emission at its peak predicted rate, well below the pile-up limit for a 30 s frame time.

### 4.1.6 Calibration

The MCP optic and detector calibration will first be determined in a pre-launch ground calibration campaign. The 27m beamline within the Tunnel Test Facility (TTF) at Leicester University would be used to characterise the individual units of the optic and detector plane prior to assembly. The end-to-end calibration of the complete instrument could be performed at the Panter facility[3] that has a 130m beamline and has been used for many previous instruments such as XMM-Newton EPIC and the Swift X-Ray Telescope. In-orbit calibration of the CCD gain and charge transfer efficiency (CTE) will be based on data from a standard onboard Fe$^{55}$ source, which will continuously illuminate the focal plane with line emission at Mn K$_\alpha$ and K$_\beta$ energies (5.9 and 6.5 keV), outside the range of the line emission from SWCX.

In addition, we will draw on the calibration heritage provided by the current generation of X-ray instruments[4] with a programme of regular observations of suitable standard candle astrophysical sources that have well-characterised X-ray spectra, such as line-rich supernova remnants.

### 4.1.7 Current heritage

The X-ray WFI draws on decades of development work on X-ray detectors for astrophysical applications. No design activity is required with regards to the baseline CCD format or packaging. These devices are already being manufactured for ground-based astronomy applications. For X-ray astronomy, processing the reverse side (with respect to the electrode structure) of the CCD to thin the passive silicon layer is required to achieve acceptable quantum efficiency at soft X-ray energies. This is an established industrial process.

Space qualified readout electronics will be available for the device since a derivative used for the SUVI instrument on NASA/NOAA's GOES-R mission has undergone full qualification.

The components of the optic array can currently be considered to be at an advanced stage of technology readiness, based on the environmental testing of the BepiColombo MIXS-C structural and thermal model, whose development has been funded by ESA.

### 4.1.8 Critical issues

Primary critical issues are avoiding damage to the CCDs during operations by excessive optical loading, and potential damage to the optic array by thermal loading. Both can be avoided by keeping the Sun outside of the stray-light path to the optic. The minimum geometrical Sun avoidance angle is from 28° to 39° depending on the roll angle of the spacecraft in its current configuration. There is sufficient contingency within the mass and volume budget of the mission profile to allow refinement of the design of the baffle (all enclosed in the spacecraft body; see Fig. 14), to provide additional shading from the bright Earth, which will be outside of the detector FOV but has a stray-light path to the optic.

---

[3] http://www.mpe.de/panter/about_en.html

[4] http://www.iachec.org

Radiation damage to the CCDs is not expected to be a critical issue. Damage from micrometeorite strikes is likely to be a very low probability event. XMM-Newton has suffered from only four such events in 10 years of operations and has a geometric optic area 13.4 times that of the WFI.

## 4.2 Plasma package: Proton-Alpha Sensor (PAS), Ion Composition Analyser (ICA)

### 4.2.1 Measurement technique

A plasma package is included on AXIOM to measure full 3D ion velocity distribution functions and thus derive basic solar wind parameters (e.g. density, velocity, temperature). These in situ measurements are essential for monitoring the conditions of the solar wind (by sampling protons and alpha particles), and characterising the heavy ions input which drives the charge exchange processes leading to X-ray production (thus an ion composition analyser is included, optimised to detect high charge state ions, e.g. $C^{6+}$, $C^{5+}$, $N^{7+}$, $N^{6+}$, $O^{8+}$, $O^{7+}$, $Fe^{18+}$, $Fe^{17+}$, $Mg^{12+}$, $Mg^{11+}$, etc.). As such, the required sampling cadence (order of minutes) is much lower than that normally employed in analysers dedicated to specific solar wind investigations.

### 4.2.2 Instrument conceptual design and key characteristics

Plasma packages of the kind required by AXIOM are currently flying on ACE and STEREO and are also under development for Solar Orbiter. The particle instruments described here are primarily based on the Solar Orbiter Solar Wind Analyser (SWA) package, for which UCL/MSSL is the PI institute, optimised to address the AXIOM scientific requirements, which are overall less severe than those of the SWA. The two analysers in the plasma package share a common Digital Processing Unit (DPU).

**Proton-Alpha Sensor (PAS)**: This consists of an electrostatic analyser (EA) with an ion steering system designed to measure the full 3D velocity distribution functions of the major solar wind species, protons and alpha particles, in the energy range 0.2 – 20 keV/q, q being the particle charge (see Fig. 12).

Arriving solar wind ions enter the instrument through an exterior aperture grid. The electrostatic steerers use voltages applied to either the upper or lower deflector electrodes as needed to steer ions from a desired arrival direction into the energy analysis section. The steerers acceptance angle can be varied through the required ±15° range (which takes into account the fluctuations in the solar wind direction and the orbital plane orientation with respect to the Sun). The required performance parameters of the PAS are listed in Table 4.

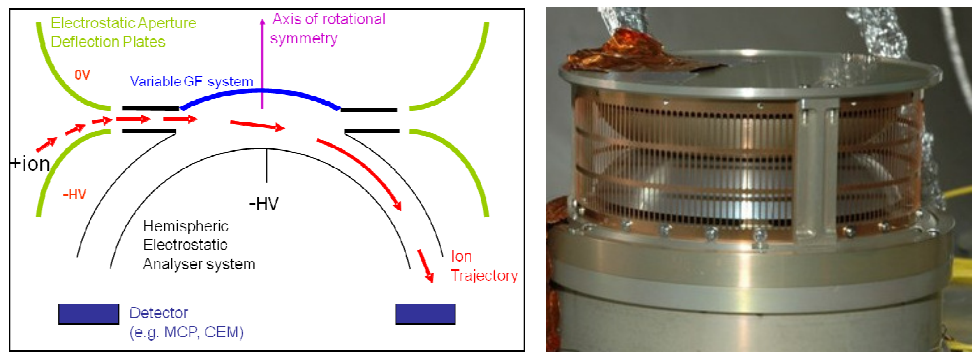

**Fig. 12** Left: Schematic of the working principle of a top-hat analyser suitable for implementation as the PAS. Right: Photo of the prototype developed at UCL/MSSL incorporating this design.

A significant enhancement on previous versions of the instrument could be the addition of a novel 'top-cap' electrode, providing an electrostatic variable geometric factor system. The application of a potential to the top-cap reduces the geometric factor of the system by restricting the fluxes of particles that pass through the electrostatic steerers before they enter the energy analysis section. The inclusion of a variable geometric factor system would permit the PAS instrument to be operated in an electron detection mode (by reversal of the polarities of the applied voltages) as the instrument sensitivity required for electrons is considerably higher. Electron measurements would add detail to the characterisation of solar wind conditions, but are not a requirement.

**Ion Composition Analyser (ICA)**: This comprises an EA module with ion steering coupled with a time-of-flight (TOF) telescope with solid state detectors (SSDs) for total ion energy measurements. ICA will measure five key properties for all ions: charge (q), mass in the range 2 – 56 amu/q, energy (E) and direction of incidence ($\theta$, $\varphi$).

Solar wind ions enter a small aperture and are steered into a top-hat (TH) analyser that aligns their trajectories with the entrance of the EA. This selects ions within the appropriate E/q range. The analyser with the ion steerer thus provides E/q and elevation. The energy (E/q) resolution of the analyser is 5.6% and the elevation angle resolution is 3°. The TOF-SSD telescope, including its post-acceleration voltage, is designed to provide measurements of azimuth, time-of-flight and total energy.

After passage through the steerer and analyser, the ions converge at a focal plane that is co-aligned with the grounded grid that separates the energy analysis section from the TOF-SSD telescope. After sufficient stand-off of ~1 cm to allow safe operation at the highest voltages, a segmented ultra-thin (~1 μg cm$^{-2}$) carbon foil emits electrons as the accelerated ions cross the foil and undergo straggling. The electrons are deflected onto a MCP and provide the START signal for the TOF analysis, as well as the azimuth position.

The ion traverses approximately 10 cm in a field-free volume, before hitting the SSD array and emitting another set of electrons which are deflected onto a STOP MCP, completing the TOF measurement. Each SSD comprises 16 pixels, spanning 110° in azimuth. Three SSDs cover practically the full 360° FOV, and also accommodate azimuth scattering from heavy ions as they pass through the carbon foil. The angle, TOF, and energy resolutions of the TOF-SSD combination are sufficient to satisfy the angle and mass resolution requirements.

### 4.2.3 Performance assessment

Table 4 shows an indicative list of the performance parameters of the AXIOM plasma package, based on the Solar Orbiter SWA currently in Phase A/B. The parameters will be optimised in line with AXIOM requirements, which are overall less severe than those of the SWA. One such optimisation will concern the low energy threshold of the ICA, which, as quoted in Table 4 (0.5 keV/q), is below the solar wind proton peak (typically around 1 keV/q). To avoid saturating the TOF one solution (adopted by the Solar Orbiter SWA) could be for the EA to sweep down in energy to the level at which the proton peak appears, and then reverse the sweep. Hence the instrument threshold might technically be lower, but the lowest part of the energy range would not be routinely sampled.

**Table 4  Performance summary of plasma analysers**

| Parameter | Range/ resolution | PAS | ICA |
|---|---|---|---|
| Sensors | | 1 x EA | 1 x EA, 1 x TOF-SSD |
| Mass | Species | H, He | C, N, O, Fe, Mg, etc. |
| | Resolution ($m/\Delta m$) | - | 5 |
| Energy | Range | 0.2 – 20 keV/q | 0.5 – 100 keV/q (AZ) 0.5 – 16 keV/q (EL) |
| | Resolution ($\Delta E/E$) | 7.5% | 5.6% |
| Angle | Range (AZ) | 360° | 360° |
| | Range (EL) | ± 15º | ± 15° |
| | Resolution (AZ× EL) | < 2° | < 2° |
| Temporal | Resolution | 3 s | 5 min |
| Geometric factor | Per pixel (cm$^2$ sr eV/eV) | 4 x 10$^{-5}$ | 2 x 10$^{-5}$ |

### 4.2.4 Pointing requirements and configuration needs

In order to obtain full 3D measurements of the solar wind ions, a key requirement for the plasma instruments is that their FOVs cover the sunward direction. This imposes a number of restrictions for instrument accommodation and imposes requirements which are strongly influenced by the mission orbital configuration.

The baseline configuration is for AXIOM to be at the Earth - Moon L1 point and implies that the spacecraft will travel approximately in the Moon's orbital plane around the Earth. Since the plasma instruments have an instantaneous FOV of 360° x 3°, by aligning the 360° FOV plane parallel to AXIOM's orbital plane, the instruments will be able to capture solar wind ions through the complete spacecraft orbit around the Earth. In addition, using the ±15° ion steering capability, the instruments will be operated to cover the required FOV beyond the instantaneous 3°, compensating for solar wind direction fluctuations and orbit inclination, as mentioned above.

An additional requirement is that the 360° FOV is un-obstructed by the spacecraft (and propulsion module) body. All of this suggests mounting the plasma package on the top side (where the solar panels are located), or the bottom side of the hexagonal spacecraft (see Fig. 14). To avoid obstructing the FOV, the plasma instruments have to protrude out of the spacecraft by between 25 cm and 1 m depending on where they are mounted; since this may exceed the envelope of the Vega fairing, the current plan is to mount the plasma package on a boom of 1 m length (worst case), which would be deployed after launch.

**Combined sensors (PEPE) option**: The mass and power resources for the plasma package can be significantly reduced by adopting a combined EA capable of performing both the fast major species ion sampling and the relatively slower but detailed composition measurements. An example of such a sensor that could be modified and exploited is the PEPE (Plasma Experiment for Planetary Exploration, Young et al. 2007) charged-particle spectrometer, flown onboard the Deep Space 1 mission, capable of simultaneously measuring and resolving the velocity distribution of electrons and ions and their mass composition. Modifying the electron analyser to detect ions (usually involving simple reversal of the voltage polarities) and optimising the instrument parameters for fast measurements of the major ion species would satisfy AXIOM's requirements and provide significant resource savings. As an example, the estimated mass for a combined instrument is 6.0 kg (compared with 9.0 kg for individual sensors).

### 4.2.5 Operating modes

Both PAS and ICA instruments will operate continuously in 'normal' mode, sampling at 3 s and 5 min cadence for protons/alpha particles and ions respectively (there is no requirement for a fast, or 'burst' mode, since AXIOM will only monitor the solar wind and not study it in detail). During CMEs the ICA setup could be optimised to select and study specific ion species.

### 4.2.6 Calibration

The AXIOM plasma package will be calibrated on the ground and in orbit. The evolution of the MCP detector gain as a function of applied high voltage can be monitored in orbit using periodic tests. Internal calibration of ICA will include validating instrument efficiencies and comparing its response to various ions. Cross-calibration between PAS and ICA is envisaged in order to refine the geometric factor knowledge for the sensors, and to verify the ground calibration. Measurements will also be compared with the instrument forward models.

### 4.2.7 Current heritage

UCL/MSSL have very substantial heritage of developing and operating plasma instrumentation in space, over several decades and on many space missions. The components of the proposed analysers are deemed to be already at an advanced level of technology readiness on the basis of the Solar Orbiter development which has reached Phase A/B stage.

### 4.2.8 Critical issues

The procurement of PAS and ICA does not present challenges above and beyond what is already planned for the Solar Orbiter SWA.

## 4.3 Magnetometer (MAG)

### 4.3.1 Measurement technique

The scientific goal of the magnetometer experiment (MAG) is to establish the orientation and magnitude of the solar wind magnetic field. This information is of critical importance for understanding and interpreting the X-ray imaging data. The magnetometer will also be used in combination with the in situ plasma measurements to detect interplanetary shocks and solar wind discontinuities passing over the spacecraft. These interfaces, moving at hundreds of km s$^{-1}$ in the solar wind, take only a few seconds to pass over the spacecraft. To fully characterise the properties of these boundaries in the solar wind, it is often necessary to examine their substructure (for example by using analysis techniques to determine boundary orientation); in a baseline design, the magnetometer would therefore measure the magnetic field at a sampling rate of up to 32 Hz.

### 4.3.2 Instrument conceptual design and key characteristics

To accurately measure the solar wind magnetic field strength and orientation, it is necessary to separate the ambient field from magnetic disturbances created by the spacecraft. To do this, the first step is to ensure that the spacecraft is as magnetically clean as possible. The second step is to remove the magnetometer sensors from the spacecraft body by mounting them on a spacecraft-provided rigid boom, with an electronics unit on the main body of the spacecraft. Two sensors, mounted at different distances from the spacecraft, should be used. This combination of two sensors allows the instrument to operate as a gradiometer and enables the background spacecraft magnetic field to be accurately subtracted from the measurements.

A baseline design for MAG is a dual redundant digital fluxgate magnetometer consisting of two tri-axial fluxgate sensors connected by harness to a spacecraft-mounted electronics box. The fluxgate sensor design is known to exhibit good stability and has extensive space heritage. In this baseline design, the electronics unit consists of an instrument controller unit, a power converter unit, and dedicated front-end electronics for each of the two magnetometer sensors. An example of a fluxgate sensor is shown in Fig. 13.

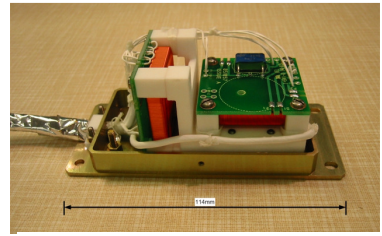

**Fig. 13** Photograph of a fluxgate sensor

### 4.3.3 Performance assessment

Fluxgate magnetometers have been used on multiple satellite missions to measure the properties of the solar wind magnetic field at 1 AU; as such the performance of the instrument is capable of achieving the science objectives with low risk**.**

### 4.3.4 Configuration needs

The optimum placement of the magnetometer sensors will depend on the estimated strength of the spacecraft magnetic field, and the underlying magnetic cleanliness programme, but, based on previous experience, the two sensors should be separated by approximately 2 m and the inboard sensor should be at least 1 m from the spacecraft body.

### 4.3.5 Operating modes

MAG will be able to operate with different sampling rates (normally at 16 Hz, given the low cadence of the X-ray imaging, but with the possibility of reaching up to 32 Hz, as explained above) and different ranges for the strength of the field depending on the ultimate mission requirements.

### 4.3.6 Calibration

MAG will be calibrated on the ground to establish e.g. sensor offsets, orientation of the sensors relative to the spacecraft, and temperature dependence of electronics and sensor; also it will be system-tested in artificial fields in a laboratory coil facility. Calibration in orbit will include measurement of sensor noise, removal of slow variation in stray spacecraft fields, by comparing data with the expected theoretical variations in solar wind magnetic field fluctuations, and removal of fast variations in stray spacecraft fields, by using the gradiometer analysis to compare the measurements at the two sensors.

### *4.3.7 Current heritage*

The fluxgate sensor has strong space heritage and a high level of technology readiness. It has successfully flown on Cassini and Double Star (note that in the operation of the magnetometer on Double Star Tc1 considerable experience was gained in gradiometer analysis) and is part of the selected payload for Solar Orbiter. The front-end electronics will be based on a new digital implementation for Solar Orbiter. The instrument controller and power converter will be designed specifically for this particular implementation but would feature extensive design heritage from instruments flying on the Cluster, Cassini, Rosetta, Double Star, Venus Express, Bepi-Colombo and Solar Orbiter spacecraft.

### *4.3.8 Critical issues*

Because of the necessity to operate the magnetometer as far as possible in a magnetically clean environment, a magnetic cleanliness programme must be established from the beginning of the project, so that cleanliness requirements are taken into account from the start of the spacecraft design phase. For example, use of magnetic materials should be controlled as far as possible. Accommodating the magnetometer on a boom (and examining carefully the achievable boom length) is part of the strategy, so as to remove it as much as possible from spacecraft generated DC fields (which should be kept below 10 nT at the outboard sensor).

Changes in the spacecraft magnetic field can normally be deconvolved by calibration as long as they are below 0.5 nT and 1 nT for slow and fast changes respectively. These are challenging requirements, so magnetic cleanliness may become a driver; further investigation will be needed, drawing on the substantial experience built up over several past and present space missions.

# 5 System requirements and spacecraft key issues

As the baseline, AXIOM consists of a science spacecraft attached to a propulsion module (possibly a re-use of that of LISA Pathfinder). An alternative architecture of a fully integrated spacecraft system is also possible, but would retain much dead mass once the operational orbit is reached, increasing propellant and power consumption for pointing and station-keeping.

## 5.1 Attitude and orbit control

A 3-axis stabilised spacecraft is required for the AXIOM mission to enable the X-ray telescope to observe the Earth's magnetosphere continuously. The pointing requirements for the X-ray telescope (15 arcmin accuracy) are not overly stringent, and can be achieved with four reaction wheels for pointing and thrusters for reaction wheel off-loading. Two star trackers and two Sun sensors provide attitude and orbit measurement. The Sun sensors are also used for Safe Mode attitude sensing.

## 5.2 Missions operations

As mentioned in sec. 3, AXIOM will be launched on a Vega rocket into a highly elliptical orbit. Ten or so apogee raising manoeuvres, taking place over less than a month, will be followed by a perigee raising trajectory using Weak Stability Boundary effects, taking around 4 to 6 months. Deployment of the two instrument booms could take place at this point, along with commissioning of the instruments. Following final capture at the Moon and manipulation into the Lissajous orbit around the L1 point the propulsion module can be separated. Once nominal operations configuration is reached, the AXIOM spacecraft will track the region of interest – the magnetopause on the Sun-side of the Earth – while orbiting both within the 14 day Lissajous orbit about the L1 point, and in the greater 28 day L1 orbit around the Earth (see Fig. 5). The minimum Sun angle that can be withstood by the WFI will determine the duration of the off-pointing period, when the spacecraft will need to rotate in attitude to point away from the Sun. Secondary science can be performed at this time.

## 5.3 Spacecraft configuration and resource requirements

The AXIOM spacecraft configuration is driven by the three instruments and the solar arrays. A simple hexagonal box is selected for the platform to maximise the use of the Vega fairing volume and provide sufficient mounting for all units. The overall configuration is shown in Fig. 14.

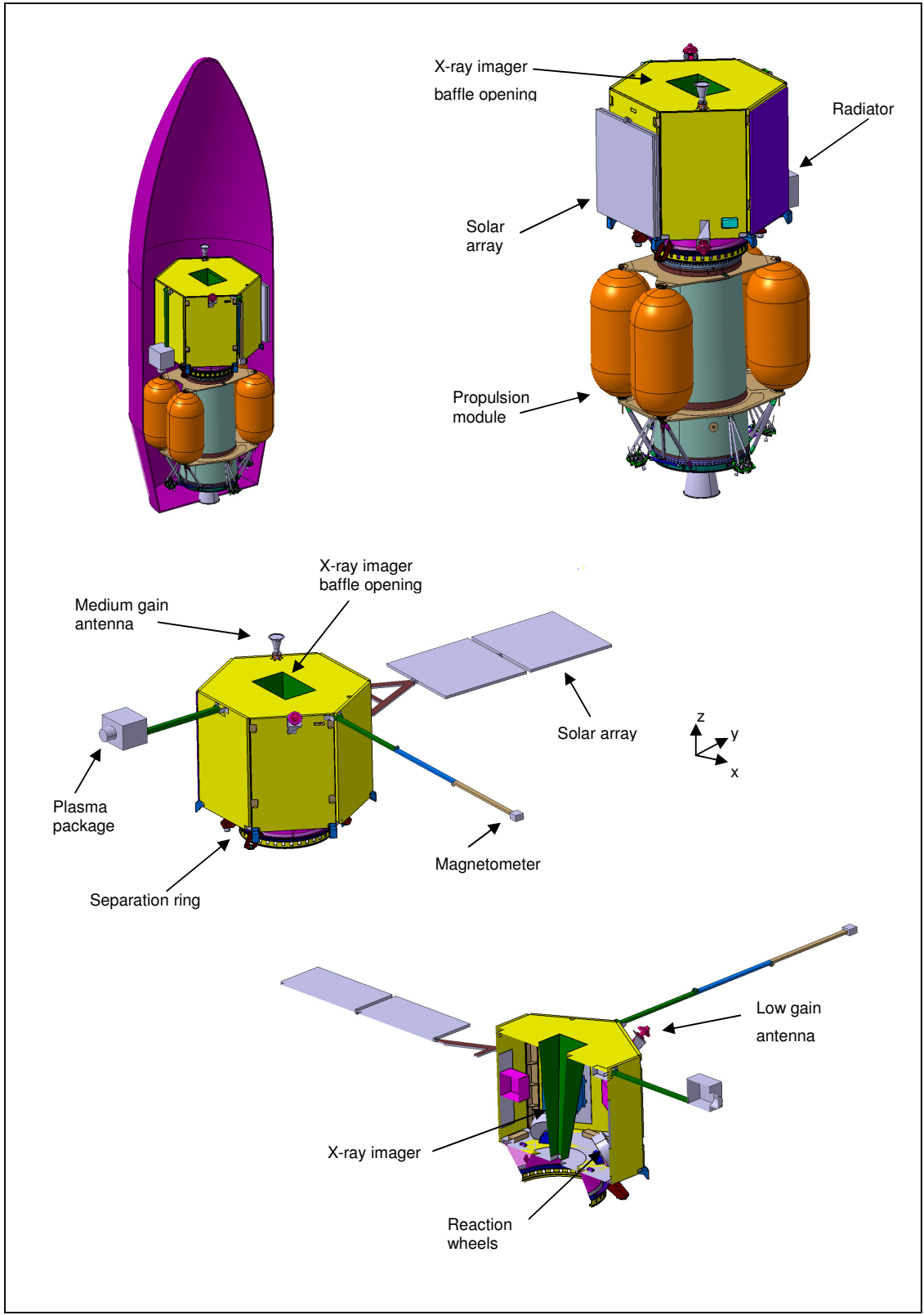

**Fig. 14** Stowed AXIOM configuration in Vega fairing (top left) and after fairing jettison (top right). Deployed AXIOM science spacecraft (middle left) and internal view (bottom right).

The X-ray telescope is located in the centre of the spacecraft to provide structural support and additional radiation shielding, and to ensure the centre of gravity is minimised during launch. Extra length is available for a baffle in this configuration, to reduce stray-light effects, and the configuration presented has a telescope length of 70 cm and a baffle length of 55 cm (reducing the Sun constraint to $< 40^\circ$). Aluminium-equivalent shielding of 1.5 mm is likely to be required (see sec. 5.4). The solar arrays and plasma package must be Sun facing at all times, driving the accommodation of each to the N and S faces (+Y and –Y directions) respectively. In this position, the solar array can use its drive mechanism to continuously rotate to track the Sun, with only one degree of freedom rotation required. The plasma package has a 360° FOV, and at the end of the boom is able to have a continuous view of the solar wind stream. The magnetometer is currently required to be on a 3 m boom, and is pictured in the +X direction.

Two low gain antennas are accommodated at diametrically opposite positions, to ensure a full $4\pi$ sr coverage. A medium gain antenna is pointing in the same direction as the X-ray telescope, so that the Earth is always within the FOV of the antenna. The solar array size, of approximately 2 $m^2$, will produce the required power.

Table 5 summarises the resource requirements for the payload and the spacecraft.

**Table 5  Resource requirements for AXIOM payload and spacecraft**

| Resources table | WFI | PAS[a] | ICA[a] | MAG | Spacecraft | Total including all margins |
|---|---|---|---|---|---|---|
| **Mass (kg)** | 30 | 4 | 9 | 3.2 | 254 (dry) 19 (fuel) | 434 (4% below launch capability) |
| **Power (W)** | 30 | 5.5 | 8.5 | 2.5 | 357 | 568 |
| **Telemetry (kbit/s)** | < 100 | < 14 | | < 3.2 | 8 | 125 |

[a] Including common DPU

## 5.4 Environmental constraints

The AXIOM radiation environment is similar, although slightly higher, to that of LISA Pathfinder. A 1.5 mm aluminium-equivalent shielding will be sufficient for most units, with the spacecraft receiving a total mission ionising dose of 100 krad with this shielding level. The dose can be reduced to 30 krad with spot shielding of 3 mm. This may be necessary for the WFI CCDs and the plasma package MCPs, which are the critical elements with respect to radiation levels.

The Earth – Moon L1 point has no severe thermal requirements, as the orbit will be approximately 62,000 km from the Moon and 323,000 km from the Earth. At this distance, the view factor of both the Moon and Earth is very small, so no accounting for albedo and infrared radiation is required. The spacecraft is effectively in deep space at 1 AU, and will only encounter very short eclipse periods.

The magnetic cleanliness requirements are likely to be a driver for the AXIOM mission, as described in section 4.3.8. Electrostatic cleanliness is also required to avoid spacecraft charging.

## 5.5 Current heritage and critical issues

All AXIOM spacecraft units are based on existing designs with significant heritage and most of them are at a very high level of technology readiness.

The major requirement for the AXIOM spacecraft is the view of the Sun required for the solar arrays and the plasma analyser package. This drives the accommodation of both to the North-South direction, impacting on the remaining configuration.

# 6 Science operations and archiving

The AXIOM X-ray WFI will point at the nose or the flanks of the Earth's magnetosphere continuously, as long as the Earth or/and Sun constraints allow it; at the same time the plasma instruments and the magnetometer will monitor the solar wind in order to derive its basic parameters and set the X-ray observations into context. At times when the front of the magnetosphere is not viewable, the WFI will carry out secondary science, such as accumulate long exposures of the near-Earth magnetotail or observing comets. Alternatively, the WFI performance will be calibrated by making observations of astrophysical sources such as supernova remnants. Variability in the data monitoring the Earth's magnetosphere response to variations in solar wind characteristics will be rather slow (typically on timescales of a few minutes or longer), so the instruments will operate in the same modes practically all the time, without the need for changing operational modes on-the-fly.

Because of the monitoring character of the AXIOM mission, and of the limited range of observing modes, planning the observations is expected to be a relatively routine task, except for the periods when the WFI cannot observe the nose of the magnetosphere, and alternative targets will be sought.

The archived database of AXIOM observations will be a novel and unique resource providing a global view and a measure of the response of the Earth' magnetosphere under the influx of the solar wind, and especially of its response to changes in the wind conditions. As such it will constitute a golden reference data bank for validation of solar-terrestrial interaction models and understanding space weather effects.

# 7 Conclusions

AXIOM is a novel, high science and low cost concept mission; it has the potential to revolutionise magnetospheric physics by providing images and movies of the dynamic solar wind – magnetosphere interaction based on SWCX X-ray emission, using state-of-the-art detection techniques. This capability will also overcome the key obstacle to effective communication of solar-terrestrial physics to the general public, in that it involves the study of processes which are, except for the aurora and the solar corona during eclipse, invisible to the naked eye. By providing an 'X-ray' of the magnetosphere AXIOM constitutes a natural hook through which the general public of all ages can engage with space science and our home planet's space environment.

AXIOM aims to tackle in a new way the fundamental issues raised by ESA's Cosmic Vision quest to establish *'How does the Solar System work?'*. Unlike the local measurements made by the majority of previous and current missions, AXIOM looks at the entire magnetic environment of the Earth as the target, embedded in the solar wind flow. Ultimately, a better understanding of how energy is transferred from the solar wind into this environment will have crucial implications on our modelling and eventually forecasting of space weather effects.

# Acknowledgements

We thank Rafael Guerra Paz for creating the AXIOM logo, and the anonymous referee for suggestions that have helped clarify and add content to the manuscript.